\documentclass[10pt,twocolumn,pra]{revtex4-1}
\usepackage{amssymb}
\usepackage{graphicx}
\usepackage{bm}
\pdfoutput=1
\begin{document}

\title{Phase diagram of Landau-Zener phenomena in coupled one-dimensional Bose quantum fluids}
\author{Santiago  F. Caballero-Ben\'\i tez$^{1,2}$ and Rosario Paredes$^1$,} 
\address{$^{1}$ 
 Instituto de F\'{\i}sica, Universidad
Nacional Aut\'onoma de M\'exico, Apdo. Postal 20-364, M\'exico D.
F. 01000, M\'exico. }
\address{
$^{2}$Nonlinear Physics Centre, Research School of Physical Sciences and Engineering, Australian National 
University, Canberra ACT 0200, Australia.}


\date{\today}
\begin{abstract}

We study stationary and dynamical properties of the many-body Landau-Zener dynamics of a Bose quantum fluid confined in two coupled one-dimensional chains, using a many-body generalization recently reported [Y.-A. Chen et al.], within  the decoupling approximation and the one-level band scheme. The energy spectrum evidences the structure of the avoided level crossings as a function of the on-site inter particle interaction strength. On the dynamical side, a phase diagram of the transfer efficiency across ground-state and inverse sweeps is presented. A totally different scenario with respect to the original single-particle Landau-Zener scheme is found for ground-state sweeps, in which a breakdown of the adiabatic region emerges as the sweep rate decreases. On the contrary, the transfer efficiency across inverse sweeps reveals consistent results with the single-particle Landau-Zener predictions. In the strong coupling regime, we find that there is a critical value of the on-site interaction for which the transfer of particles starts to vanish independently of the sweep rate. Our results are in qualitative agreement with those of the experimental counterpart.
\end{abstract}
\maketitle

\section{Introduction.}\label{Intro}

The original Landau-Zener (LZ) problem involves a two-level system whose energy separation varies as a linear function of time. These two states can be associated to the lowest states of a single-particle in a double-well potential.  As it is well known  \cite{Landau,Zener}, the transition probability  between the two energy states is the well known LZ formula  $P_{LZ}=\exp(-2 \pi J^2/\hbar \alpha)$  written in terms of the tunnelling-coupling parameter $J$ and the time-dependent detunning $\Delta =\alpha t$, accounting for the probability that a particle initially in the bottom well at $t=-\infty$ reaches at $t=\infty$ the opposite well. Recent experiments have demonstrated the many body generalization of the LZ phenomena by loading an ultracold Bose gas into a pair of coupled one-dimensional chains where a controllable  inter-chain sweep leads to the observation of both, avoided crossings and breakdown of adiabatic inter-chain transfer \cite{IBloch}.

Before the experimental realization of the many body LZ generalization, several theoretical descriptions of a Bose-Einstein condensate (BEC) confined in a double-well potential have addressed the LZ problem within two intrinsically different approaches. We make reference to the two-mode mean-field and Bose-Hubbard schemes inherited from the Gross-Pitaevskii and full quantum approaches \cite{Zobay,Wu,Wu2,Witthaut,Liu,Jona-Lasinio}. As a result of incorporating the linear variation in time between the two levels, all of those treatments suggested a breakdown of the adiabatic limit, that is, that the LZ transition probability does not vanish even in the adiabatic limit. For example, the non-linear two-level system shows that the mean-field interactions among particles tend to increase the tunnelling probability and that there exists a critical value of the interaction strength beyond which the transition probability becomes nonzero even in the adiabatic limit \cite{Wu}. On the other hand, a stationary phase approximation leads to a characteristic scaling or power law for the critical behavior that occurs as the nonlinear parameter equals the gap of avoided crossing energy levels \cite{Liu}. Regarding the asymmetric LZ tunnelling in a periodic potential, Jona-Lasinio et. al. \cite{Jona-Lasinio} found that the tunnelling rates for the two directions of tunnelling are not the same. Tunnelling from the ground state to the excited state is enhanced by the nonlinearity, where as in the opposite direction it is suppressed. Even more, a LZ formula has been derived for the two mode many-particle scenario \cite{Witthaut}. 

Here, we are devoted to the description of the many body LZ generalization described in the first paragraph. That is, we are interested in describing a BEC confined in two-coupled 1D chains where the potential depths, defining the lattice sites, are linearly modified in time such that a sweep from an initial energy difference $-\Delta$ to a final one $\Delta$ is achieved. To analyze such a system, we shall use a model Hamiltonian that incorporates both, the full quantum frame of the Bose-Hubbard Hamiltonian \cite{Dounas-Frazer} and a decoupling approximation scheme \cite{Stoof,Fisher} that involves certain superfluid order parameters. These are $\psi_\nu=\langle b_{i,\nu}^\dagger \rangle= \langle b_{i,\nu} \rangle$, where $ b_{i,\nu}^\dagger$ and  $b_{i,\nu}$, $\nu \in \{L,R\}$ create and annihilate particles in lattice sites $i$ and side chain $\nu \in {L, R}$. The average is the expectation value in the investigated states. A previous theoretical work has employed the time-dependent density matrix renormalization group method \cite{Kasztelan} where, with a small number of links in the chains, qualitative agreement with the experiment was found. Our model simulates two coupled infinite chains with a local site Hamiltonian written in terms of the on-site non-perturbative inter-particle interaction strength $U$, the time-dependent detunning $\Delta$ and the intra- and inter-chain coupling energies  $J^{\parallel}$ and $J^{\perp}$. The model satisfactorily describes qualitatively the main features of the experiment: the transfer efficiency across ground-state sweeps, that in turn exhibits a breakdown of adiabaticity as the sweep rate decreases, and the transfer efficiency across inverse sweeps, that actually show consistent results with the single-particle Landau-Zener scenario. Regarding the single-particle LZ result the model Hamiltonian allows us to recover the result for a single dimer. In addition, since the number of particles per lattice sites in the actual experiments is not too large, it is possible to perform accurate numerical simulations of the system. Such a analysis gives rise to a richer spectrum with respect to the single dimer considered in \cite{IBloch}.

The manuscript is organized in VI sections. In section II we introduce the model Hamiltonian that provides a suitable description to the experiment. We also justify the use of the decoupling approximation and the one band scheme within the full quantum and mean-field approximation schemes. In section III we characterize the stationary states as a function of the both, the inter-particle interaction strength $U$ and the the parameter characterizing the tilt $\Delta$. In particular, we concentrate in determining the energy spectrum for the model Hamiltonian and the behaviour of the superfluid order parameters $\psi$. An exhaustive exploration of the parameter space allows us to encode the dynamical behavior of the of ground-state and inverse sweeps in a phase diagram. The latter are presented in sections IV and V respectively. Finally, in section VI a summary of the main results is given.

 \section{The model, decoupling approximation in the one-level band picture}\label{model}
 
\begin{figure}
\begin{center}
 \includegraphics[width=.35\textwidth,trim= 130 130 130 130]{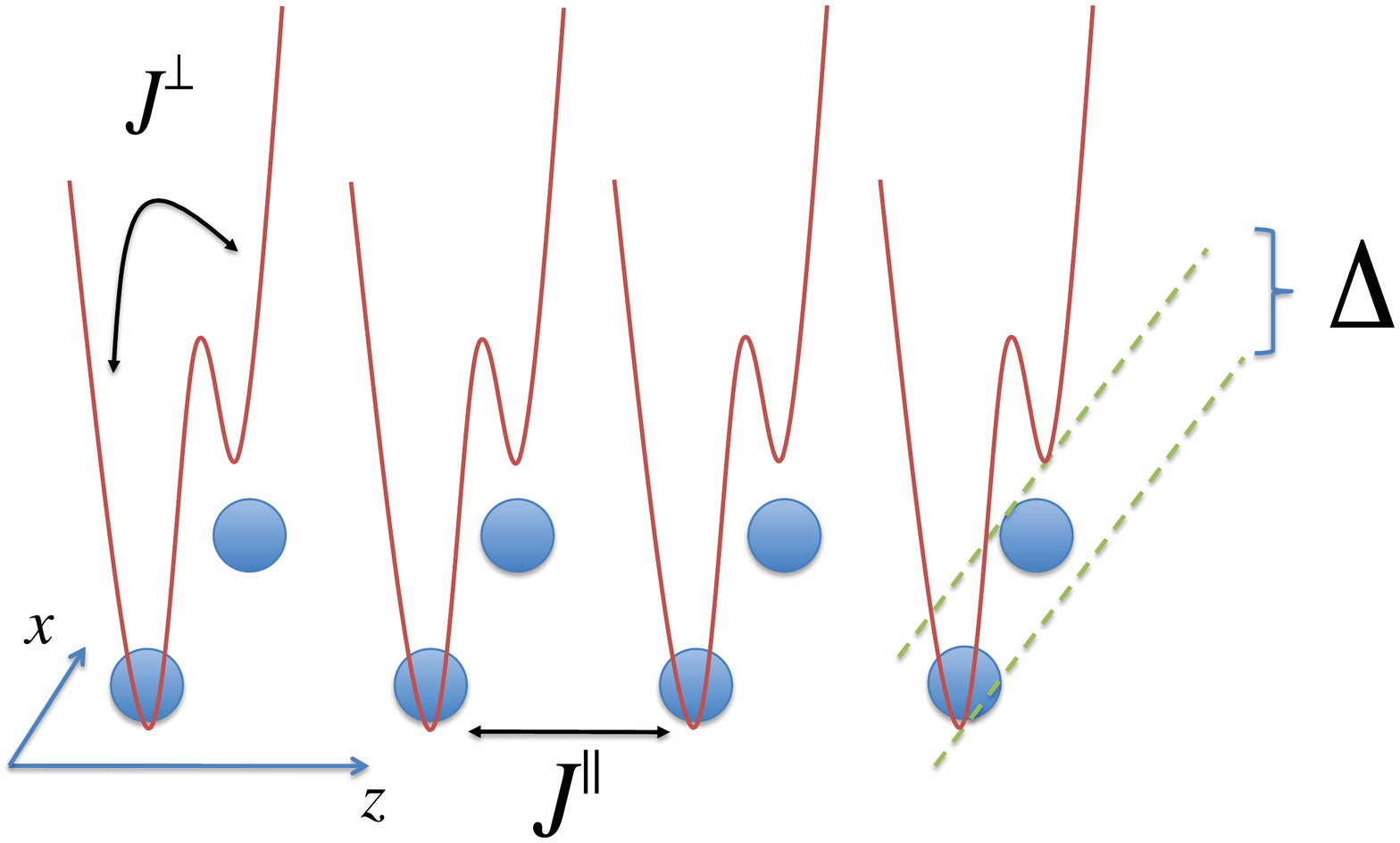}
\end{center}
\caption{(color online)
Schematic representation of the double-well trap chain potential.  The coupling energies are  intra-chain $J^{\parallel}$ ($z$ direction) and inter-chain $J^{\perp}$ ($x$ direction).  The time-dependent parameter $\Delta$  characterizes the tilt between the two wells of each site in the chain.}
\label{fig1}
\end{figure}

The physical system that we want to describe is a finite collection of interacting Bose atoms initially placed in one of the sides of a composed pairwise coupled lattice, see Fig. \ref{fig1}, and its subsequent evolution when the pairwise potential depths are linearly modified in time. As schematically shown, the transport of particles in $x$ and $z$ directions occurs as a consequence of the intrinsic tunnelling among the pairwise coupled sites, but it is also affected by the external variation of  the well potential depths with time in the form of a sweep from an initial energy difference $\Delta_i=-\Delta$ to a final one $\Delta_f = \Delta$. The parameters responsible for such a transport are the on-site inter-particle interaction strength $U$, the intra-chain coupling energy $J^{\parallel}$ ($z$ direction), the inter-chain coupling energy $J^{\perp}$ ($x$ direction) and the time-dependent parameter characterizing the tilt between the two wells of each site in the chain, namely, the detunning parameter $\Delta$. 

The purpose of the present paper is to fully characterize stationary and dynamical properties of the system described above.  To proceed, we will consider the experimental realization reported in Ref.  \cite{IBloch} where the transfer of particles in $x$ and $z$ directions is studied in what it is called the ground-state and inverse sweeps, the former accounting for a sweep starting from the ground state, namely, where all the particles initiate in the bottom well with lower energy, and the latter where the initially filled sites are the ones with higher energy. This scheme leads us to answer several questions, on one side those related to the stationary properties, such as, the energy spectrum, and  on the other side, aspects related to the dynamical behavior of the atomic population in the wells. Of special interest in this regard is the determination of the phase diagram encoding the dependence of the transfer efficiency on the parameters characterizing the dynamics. 

Let us now turn our attention to the model Hamiltonian of the system. Since the particles are not allowed to tunnel between left and right wells of {\it different} sites $i$, movements in $x$ and $z$ directions must be considered separately. Therefore, we consider the dynamics along the sites and their counterpart between the left and right chains. 

Regarding the two-well geometry and its intrinsic asymmetry supplied by the time-dependent parameter $\Delta$, we should first argue about the level picture scheme where our system is appropriately described. Due to the variation of such a parameter not just the first band but excited levels might play a significant role. As established by Dounas-Frazer et. al. \cite{Dounas-Frazer} the transport phenomena of a Bose-Einstein condensate in an asymmetric double well leads to three new energy scales with respect to the symmetric trap and one-band assumptions. These are the hopping, tilt and the energy gap between the first and second bands. The task of including at least a two-band picture (that is, four levels per site) leads 
to a Hilbert space of size $(N+1)(N+2)(N+3)/6$ with $N$ the number of atoms per site. However, since the LZ phenomena involves the presence of just two states, we should restrict ourselves to the first band of the time-varying tilted double-well potential. This can be done by appropriately taking the formal bounds where a one-band picture is well justified, which primarily assumes that the interaction energy is much smaller that the energy band difference. In the appendix we present a study of the single-particle energy levels in a double-well potential as a function of the tilt $\Delta$. To fulfill the condition that the inter-particle energy interaction is much smaller than the energy gap between the first and second bands we shall confine ourselves to values of $\Delta$ for which the bands are well separated and thus the condition with the inter-particle interaction will be considered accordingly.

Starting from the Bose-Hubbard Hamiltonian of an optical lattice \cite{IBloch}, one can write down the Hamiltonian of atoms in the two-well chain as,
\begin{eqnarray}
&&\mathcal{H}=\mathcal{H}^{\parallel}+\sum_{i}\mathcal{H}^{\perp}_i+\sum_{i}\mathcal{H}^{U}_i \nonumber
\\
&&\mathcal{H}^{\parallel}=-J^{\parallel}\sum_{\nu=R,L}\sum_{<i ,j>}\left(\hat{b}^{\dagger}_{i,\nu}\hat{b}^{\phantom{\dagger}}_{j,\nu}+\textrm{h.c,}\right) \nonumber
\\
&&\mathcal{H}^{\perp}_i=-J^{\perp}\left(\hat{b}^{\dagger}_{i,R}\hat{b}^{\phantom{\dagger}}_{i,L}+\hat{b}^{\dagger}_{i,L}\hat{b}^{\phantom{\dagger}}_{i,R}\right)+\frac{\Delta}{2}\sum_{i}\left(\hat{n}_{i,R}-\hat{n}_{i,L}\right) \nonumber
\\
&&\mathcal{H}^{U}_i=\frac{U}{2}\sum_{\nu=R,L}\hat{b}^{\dagger}_{i,\nu}\hat{b}^{\dagger}_{i,\nu}\hat{b}^{\phantom{\dagger}}_{i,\nu}\hat{b}^{\phantom{\dagger}}_{i,\nu}
\end{eqnarray}
 where the labels $\parallel$ and $\perp$ denote movements in $z$ and $x$ directions respectively.  $\mathcal{H}^{\parallel}$ represents the tunnelling along the chains, while $\mathcal{H}^{\perp}_i$ takes into account the local two-level structure at sites $i$. $\mathcal{H}^{U}_i$ corresponds to the on-site interaction energy. 
 
To consider the participation of the lattice sites, that is the transport in the $z$ direction, we use the decoupling approximation where order parameters $\psi_\nu=\langle b_{i,\nu}^\dagger \rangle= \langle b_{i,\nu} \rangle$, $\nu \in \{L,R\}$ are introduced to account for the expectation value of the tunnelling of particles between left and right wells, disregarding the position of the site $i$ within the chain. We consider  negligible the average fluctuations of creation and annihilation particle operators in sites $i$ and $j$, such that, $\Delta b_{i,\nu}^\dagger=  b_{i,\nu}^\dagger -\psi_\nu \approx 0$  and $\Delta b_{j,\nu}= b_{j,\nu}-\psi_\nu \approx 0$. This allows us to write  \cite{Stoof,Fisher,Danshita},  

\begin{equation}
\hat{b}^{\dagger}_{i,\nu}\hat{b}^{\phantom{\dagger}}_{j,\nu}\approx\psi_\nu\left(\hat{b}^{\phantom{\dagger}}_{j,\nu}+\hat{b}^{\dagger}_{i,\nu}\right)-\psi_\nu^2
\label{decoup}
\end{equation}

Thus, by substituting this expression into $\mathcal{H}^{\parallel}$ we arrive to the model Hamiltonian at each lattice site given by,

\begin{equation}
\mathcal{H}^{\textrm{loc}}_i=-J^{\parallel} \sum_{\nu=R,L}\left(\psi_\nu\left(\hat{b}^{\phantom{\dagger}}_{i,\nu}+\hat{b}^{\dagger}_{i,\nu}\right)-\psi_\nu^2\right)+\mathcal{H}^{\perp}_i+\mathcal{H}^{U}_i.
\label{Hmodel}
\end{equation}

This model Hamiltonian represents the local site energy of an infinite chain, where the interaction is non-perturbative. We should emphasize that the fact of considering a mean-field approach to separate the effective contributions by site leads us to the task of self-consistently determining the order parameters  $\psi_\nu$, $\nu \in \{L,R\}$ to specify the Hamiltonian (\ref{Hmodel}).

\begin{figure}
\begin{center}
 \includegraphics[width=.3\textwidth,trim= 100 100 100 100]{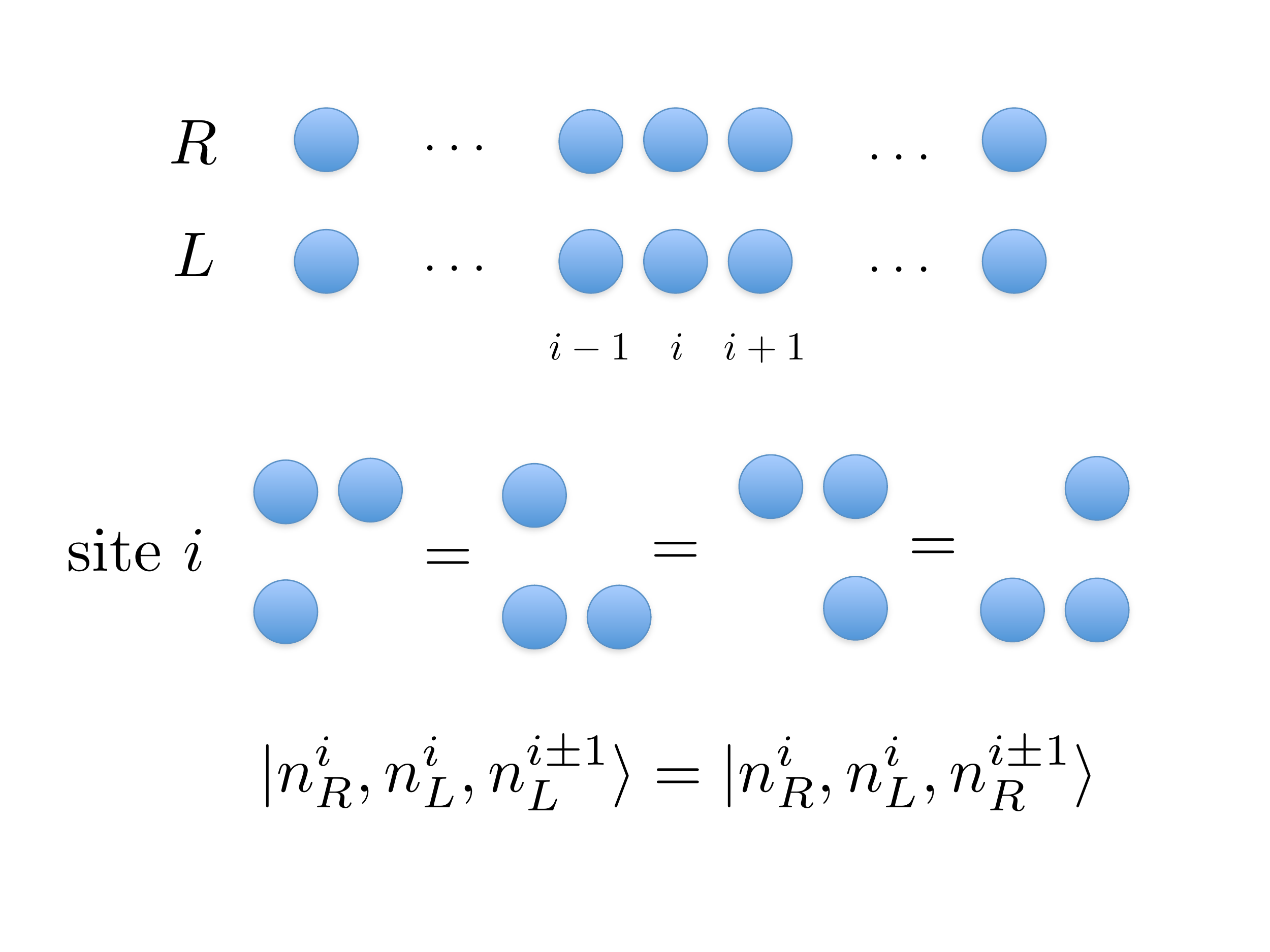}
\end{center}
\caption{(color on-line)
Schematic representation of the lattice sites in the chain of double-wells.}
\label{sites}
\end{figure}

Regarding the Hilbert space size, one can see that the Fock space of the coupled chains can be described in terms of the  occupation number of sites $i$ and $i+1$. In Fig. \ref{sites} we illustrate the equivalence among both, sites $i-(i+1)$ and $(i-1)-i$ and sides $R$ and $L$. The Fock space elements can be written as $|n_R^i,n_L^i,n_L^{i+1}\rangle$ where $n_R^i=0,...,N$, $n_L^i =0,...,N-n_R^i$ and $n_L^{i+1}=N-n_R^i-n_L^i$, being $N$ the total number of particles at each site $i$.  Thus, our model Hamiltonian, representing the local site energy of an infinite chain, corresponds effectively to $N$ particles with three ``internal" states. Consequently the Fock space at each lattice site scales with the number of particles as $\Omega = (N+1)(N+2)/2$.

It is important to notice that while the transport along $z$ direction is accounted for in a mean-field fashion, the tunnelling among left and right sides ($x$ direction) is fully quantum described. The assumption of the mean-field approach in this case is well justified in the sense that in the experimental situation near the centre of the trap, the average the number of particles per site on left and right sides, is the same along the chains. As we shall see in the next sections, this treatment allows us to study up to $N=8$ particles per site.

\section{Stationary states}\label{spectrum}

In this section we study the stationary states of Hamiltonian (\ref{Hmodel}) for a given number of particles, as a function of the interaction strength $U$ and the parameter characterizing the tilt $\Delta$. In particular, we solve the time-independent Schr\"odinger equation for the model Hamiltonian to determine the energy spectrum as a function of the parameters $U$ and $\Delta$. It is important to point out that in our calculations both, the parameters $U$ and $\Delta$  are referred to the tunnelling coupling parameter $J=J^{\parallel}=J^{\perp}$. Here and henceforth we use ${\tilde U}= U/J$ and ${\tilde \Delta}= \Delta/J$. 

To determine the entire energy spectrum of Hamiltonian (\ref{Hmodel}) we have to find the value of the order parameters $\psi_\nu$ $(\nu = L,R)$. We proceed as follows. For fixed values of $\tilde U$ and $\tilde \Delta$ we obtain by means of a variational procedure the values of $\psi_\nu$ $(\nu = L,R)$ \cite{numerics} that minimize the ground state energy $\tilde E_0(\tilde U,\tilde \Delta,\psi_L, \psi_R)$. In other words, we search for the lowest eigenvalue of the Hamiltonian $\mathcal{H}^{\textrm{loc}}_i(\tilde U,\tilde \Delta,\psi_L, \psi_R)$ by minimizing with respect to $\psi_\nu$ $(\nu = L,R)$. Once those optimal values have been determined, we calculate the entire spectrum by means of standard diagonalization methods. Then, for the same value of $\tilde U$ we repeat the same process for each value of $\tilde \Delta$

\begin{figure}
\begin{center}
\includegraphics[width=.48\textwidth]{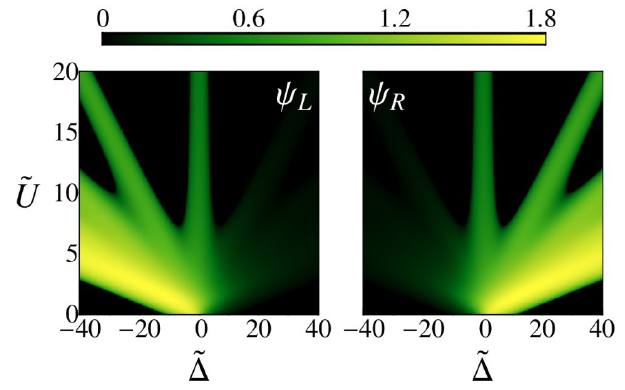}
\end{center}
\caption{(color on-line) Density plot of the optimal values of the order parameters $\psi_L$  and $\psi_R$ as  functions of the dimensionless parameters $\tilde U$ and $\tilde \Delta$ for the ground state of Hamiltonian (\ref{Hmodel}) for $N = 6$.}
\label{fig2}
\end{figure}

In Fig. \ref{fig2} we show a density plot with the optimal values of the order parameters $\psi_L$ (left) and $\psi_R$ (right), as functions of $\tilde U$ and $\tilde \Delta$. These calculations performed for $N=6$, allow us to appreciate well defined boundaries for which the optimal values of the order parameters are identically zero. These are Mott insulating (MI) state-like regions. That is, according to the identification given in section \ref{model}, $\psi_\nu=\langle b_{i,\nu}^\dagger \rangle= \langle b_{i,\nu} \rangle$, $\nu \in \{L,R\}$, the region delimited for those values of $\tilde U$ and $\tilde \Delta$ indicates that there is no tunnelling and consequently a Mott Insulator phase can be nucleated since the average fluctuations in the population per site are negligible. Conversely, a superfluid phase is associated when the order parameters are different from zero, the superfluid component in the system being maximal in the diffuse area for $\tilde U \lesssim 10$.  As expected, the indistinguishable character of  left and right wells is confirmed from the symmetry exhibited in $\psi_L$ and $\psi_R$.  
It is important to point out that the insulating regions that appear for $\tilde U\sim0$ and large $|\tilde \Delta|$ are  finite size effects similar to what was found in small Bose-Fermi systems  \cite{BF}, which in the limit of large $N$ these regions disappear.

\begin{figure}
\begin{center}
 \includegraphics[width=.48\textwidth]{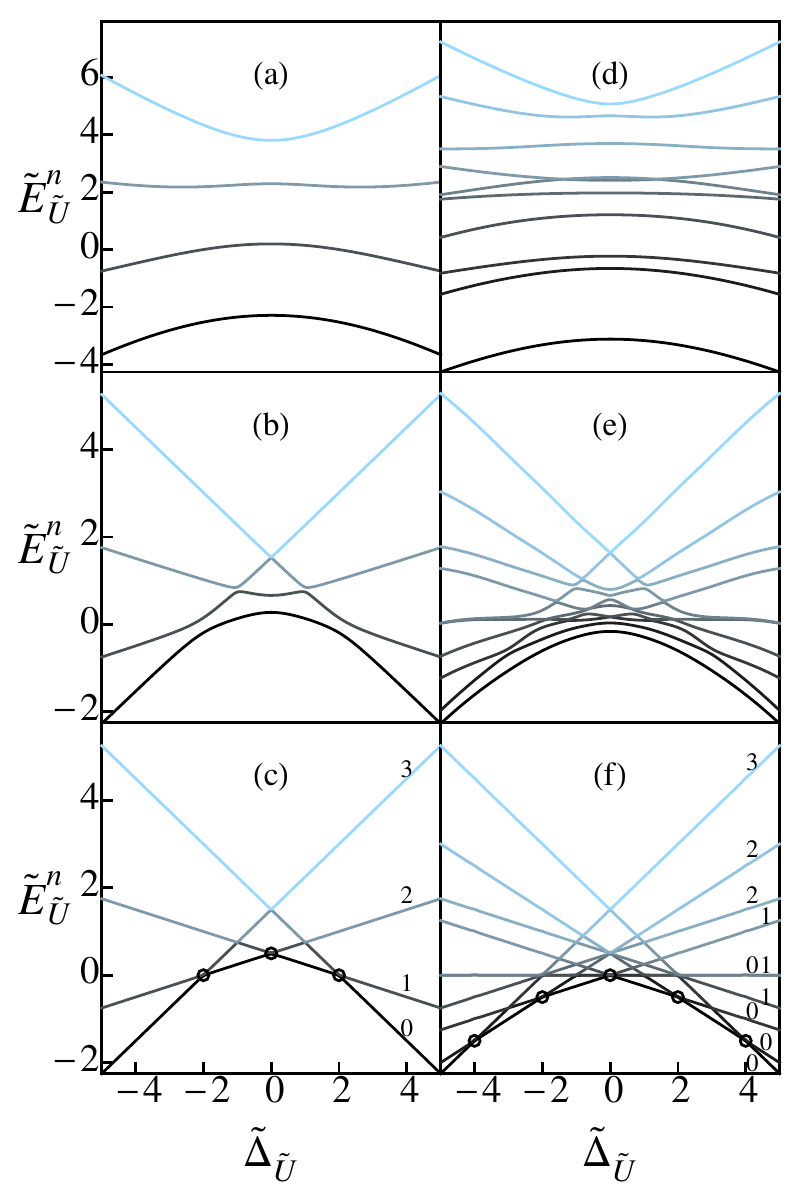}
\end{center}
\caption{(color on-line)
Energy spectrum $\tilde E^n_{\tilde U}=E_n/ J \tilde U$ of Hamiltonian (\ref{Hmodel}), for $N=3$ and $\tilde U=1\;\textrm{(a,d)},10 \;\textrm{(b,e)},100\;\textrm{(c,f)}$. The values of $\tilde \Delta$ in the $x$ axis have been rescaled with respect to the dimensionless parameter $\tilde U$, that is, $\tilde{\Delta}_{\tilde{U}}=\tilde{\Delta}/\tilde{U}$. For $\tilde U =100$ the numbers on the right refer to the number of particles in the right well, while the circles indicate the tunnelling resonances.}
\label{fig3}
\end{figure}

For illustration purposes, in Fig. \ref{fig3} we show the entire energy spectrum of Hamiltonian (\ref{Hmodel}) for $N=3$ and  different values of $\tilde U$. The values of $\tilde E_n$ and $\tilde \Delta$ have been rescaled with respect to $\tilde U$ as $\tilde{\Delta}_{\tilde{U}}=\tilde{\Delta}/\tilde{U}$ and $\tilde E^n_{\tilde U}=\tilde E_n/ \tilde U$. For $\tilde U =100$ the numbers on the right refer to the number of particles in the right well, while the circles indicate how tunnelling resonances are crossed one after another. The left figures ((a), (b) and (c)) are obtained by neglecting the transport along the $z$ axis (see Fig.1). That is, they correspond to the energy spectrum of a single double-well. The right figures of the panel contains the energy spectrum considering the transport in both, parallel and perpendicular directions. From the full panel of Fig. \ref{fig3} one can observe that although avoided level crossings are exhibited in the whole interval $[-\tilde \Delta_{\tilde U}, \tilde \Delta_{\tilde U}]$ for every considered  value of $\tilde U$, the level crossing structure is more visible as $\tilde U$ grows. As can be appreciated, the results of our model (\ref{Hmodel}) when the transport along the $z$ axis is neglected, coincide with those obtained for Bose-Hubbard Hamiltonians of one dimer only \cite{IBloch,Venumadhav}. In other words, if the on-site interaction energy $\tilde U$ between the particles dominates the tunnelling coupling $J$ ($\tilde U= U/J >> 1$), the LZ transitions split into $N$ avoided level crossings [see Fig.\ref{fig3} (c)], each corresponding to the transfer of one particle. A different situation is observed when particles can move through $x$ and $z$ axis. The LZ transition for $\tilde U= U/J >> 1$ split into $2N-1$ avoided level crossings. Regarding the number of avoided crossing and its dependence with the number of particles $N$ in the limit of non-interacting bosons, the single-particle LZ result is recovered for weaker interactions when the transport in $x$ and $z$ axis is considered.  It is important to emphasize that the fact of considering the combined transport in the parallel and perpendicular directions, in general gives rise to a richness spectrum with respect to the single dimer. Even more, leads to predict additional tunnelling resonances with respect to the single dimer.

\section{Phase diagram}\label{PD}

\begin{figure}
\begin{center}
  \includegraphics[width=.48\textwidth]{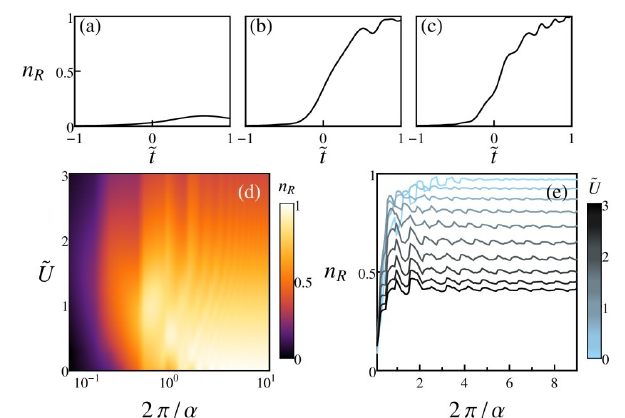}
\end{center}
\caption{(color on line) Instantaneous right evolution $n_R(t)$ from ground-state for various rates $2\pi/\alpha= 0.1 \textrm{(a)} ,1.0 \textrm{(b)} ,2.0 \textrm{(c)}$, $\tilde U=0.75$ and $N=6$. The time is rescaled with respect to  $T_{    max}$, $\tilde{t}=t/T_{    max}$, with $\tilde \Delta_0=20$. (c) Density plot of the normalized transfer efficiency $n_R$ at $t=T_{    max}$. (d) Contour plots of the normalized transfer efficiency for several values of the the dimensionless parameter $\tilde U$, see main text.}
\label{fig4}
\end{figure}

The LZ dynamics from $\tilde \Delta = -\alpha T_{  max}$ to $\tilde \Delta = \alpha T_{  max}$ for a given value of $\alpha$ are obtained by evolving in time ($\tilde \Delta=\alpha t$)  the initial state $|\Psi(0)\rangle$ associated to the ground state energy $\tilde E_0(\tilde U,\tilde \Delta)$. We have that, for a given value of $\alpha$ one should follow the evolution of the state where all the particles initiate in the left chain with $\tilde{\Delta} <0$, such that the filled chain is the one with lower energy. Experimentally, this process is identified as a ground-state sweep. The evolution is given by \cite{dt},
\begin{equation}
|\Psi(t_{i+1})\rangle\approx e^{-i H(\tilde U, \tilde \Delta)\delta t} |\Psi(t_{i})\rangle
\label{evolution}
\end{equation}
where $H$ is the local Hamiltonian (\ref{Hmodel}) and $\hbar=1$ is assumed. The initial time $t_0$ corresponds to $T_{  max}=-\tilde \Delta_0/\alpha$ while the temporal step $\delta t$ that we select, $\delta t=0.2$, produces qualitatively similar results than $\delta t \le 0.2$. It is important to emphasize that the minimization procedure described above (section \ref{spectrum}) is performed at each temporal step $\delta t$ to take into account the variation of $\tilde \Delta$ at each time interval $\delta t$. Consequently, the ground state $|\Psi(t_{i})\rangle$ is updated at each temporal step and serves as a seed for the subsequent time. The full dynamics, for a given rate $\alpha$ and inter particle interaction strength $\tilde U$, ends when the time $t_{i+1}$ reaches its maximum value $T_{  max     }=\tilde \Delta_0/\alpha$. To illustrate the evolution in time of the population on the right side, in Fig.  \ref{fig4}(a,b,c) we plot $n_R(t)$ for various rates $\alpha$, $\tilde U=0.75$ and $N=6$. The time is rescaled with respect to  $T_{    max}$, $\tilde{t}=t/T_{    max}$ with $\tilde \Delta_0=20$.

For $N=6$ we investigated the many-body dynamics for sweep rates in the interval $2\pi/\alpha \in [0.1 , 10]$ and $\tilde U \in [0,10]$ \cite{numerics2}. To condense the information of the LZ dynamics, as a function of the interaction strength ${\tilde U}$ and the sweep rate $\alpha$, we show in a phase diagram the final state that the system reaches as the initial state is evolved in the interval $[-T_{max},T_{max}]$. In Fig.  \ref{fig4}(d) we show a density plot with the normalized transfer efficiency on the right side at $t=T_{max}$. To complementary visualize the dependence of $n_R$ on $\tilde U$ in Fig.  \ref{fig4}(e) we present several contour plots of the normalized transfer efficiency as a function of the sweep rate $\alpha$, in the interval $\tilde{U}\in[0,3]$.

From Fig. \ref{fig4}(d) one can observe how the influence of both, the parameter characterizing the interaction among particles and the sweep rate $\alpha$, lead the system to well defined quantum phases from minimum to maximal values of the transfer efficiency $n_R$.  As can be observed from this figure, the effect of having $N>1$ induces the opposite behavior with respect to the original LZ phenomena when the sweep rate $\alpha$ decreases. That is, the transfer of particles does not vanish as the sweep rate tends to zero, on the contrary, when the sweep rate $\alpha$ takes smaller values, the total population initially placed on the left side is transferred almost completely  to the right chain. In other words, the main effect of having an interacting many-body system is the breakdown of adiabaticity. 

Regarding the role of the on-site interactions, one can observe that the transfer efficiency diminishes as $\tilde U$ increases. This result can be understood in the light of the well known prediction for bosonic Josephson junctions \cite{Caballero} where the transport of particles diminishes as the interaction strength increases. We observe that for a given value of the on-site inter-particle interaction the effect of decreasing the sweep rate $\alpha$, is turned from detrimental into favourable for the transfer efficiency in the $N-$body LZ scheme [see Fig. \ref {fig4} (d) and (e) ].

\begin{figure}
\begin{center}
 \includegraphics[width=.48\textwidth]{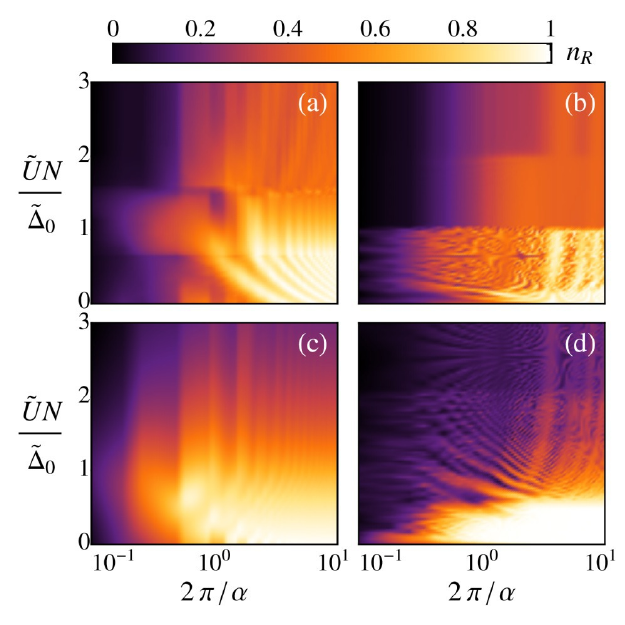}
\end{center}
\caption{(color on-line)Density plot of the normalized transfer efficiency $n_R$ of ground-state sweeps as a function of dimensionless parameters $\tilde U$ and $\alpha$ at $t=T_{ max}=\tilde \Delta_0/\alpha$ for $N=2$ (a,b) and, $N=8$ (c,d). $-\tilde \Delta_0$ is the initial value of the tilt. Parameters are:  $\tilde \Delta_0=10$  (a,c) and $\tilde \Delta_0=100$ (b,d).}
\label{fig5}
\end{figure}

To illustrate the dependence of the many-body dynamics on the initial value of the tilt $-\tilde \Delta_0$ and on the number of particles, we performed calculations for $\tilde \Delta_0=$ 10 and 100 and $N=2$ and $8$ particles per site. The results are shown in the panel of Fig. \ref{fig5}. As above, we plot the normalized transfer efficiency in a density plot for $2\pi /\alpha \in [0.1,10.0]$ and $\tilde U N/ \tilde \Delta_0 \in [0,3]$. The phase diagrams in Fig. \ref{fig5} evidence the impact of the initial tilt $-\tilde \Delta_0$ on the final state that the system reaches after a complete sweep. The region of maximum transfer is reduced for either $N=2$ and $N=8$ as $\tilde \Delta_0$ goes from 10 to 100. That is, the region exhibiting breakdown of adiabaticity becomes larger as the value of time dependent energy bias grows.  From our simulations we find that there exits a critical value of the on-site interaction $\tilde U_c\sim \tilde \Delta_0/N$ for which the atoms remain evenly distributed. Due to the relation with  the order parameters [Fig. \ref{fig2}], we find that the transfer efficiency is very sensitive with respect to the variation of $\psi_L$  and $\psi_R$. As the system evolves in time the presence and the crossing between superfluid and MI phases, give rise to reach a saturation value for the of the atoms transfer to the right side of the chain ($\max(n_R) \lesssim 0.5$ for $\tilde U \gtrsim \tilde U_c$). 

\section{Many-body dynamics: Ground-state {\it vs} inverse sweeps}

Following the experimental realization of LZ  many-body dynamics \cite{IBloch}, we extend our study to investigate the dynamics across inverse sweeps. Namely, the initial condition corresponds to $\tilde \Delta>0$ with the total particle population on the left side. The time evolution is given as before, by (\ref{evolution}), and the same minimization procedure to determine the variational order parameters $\psi_L$ and $\psi_R$ at each temporal step $\delta t$ must be carried out from $\tilde \Delta = \alpha T_{max}=\tilde\Delta_0$ to $\tilde \Delta = -\alpha T_{    max}=-\tilde\Delta_0$ for a given sweep rate $\alpha$.

 \begin{figure}
\begin{center}

\includegraphics[width=.44\textwidth]{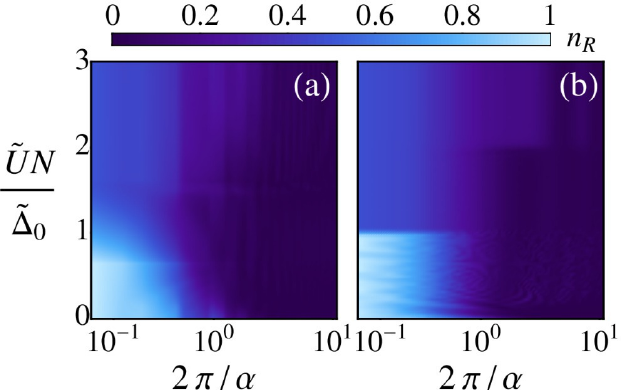}

\end{center}
\caption{(color on-line) Density plot of normalized transfer efficiency $n_R$ of inverse sweeps as a function of dimensionless parameters $\tilde U$ and $\alpha$ for $2\pi /\alpha \in [0.1,10.0]$ and $\tilde U N/ \tilde \Delta_0 \in [0,3]$. Parameters are: $\tilde \Delta_0=10$ (a),  $\tilde \Delta_0=100$ (b),  $N=2$.}
\label{fig6}
\end{figure}

To obtain the phase diagram for the inverse sweep scheme we explore sweep rates in the interval $2\pi /\alpha \in [0.1,10.0]$ for $N=2$, and set $\tilde \Delta_0=10$ and $\tilde \Delta_0=100$. For comparison purposes with the results presented in Fig. \ref{fig5}, we plot in Fig. \ref{fig6} the phase diagram of $n_R$ in terms of $\tilde U N/ \tilde \Delta_0$. In this case consistent results with the single-particle LZ scenario are found: the maximum of transfer efficiency is observed as the sweep rate increases.  The effect of increasing the value of the tilt $\tilde \Delta_0$ from 10 to 100 gives rise to a reduction of the region of maximum transfer. That is, it leads to similar qualitative features than those found for the ground state sweeps. The role of the inter particle interaction strength is perceived as in the ground-state scheme, that is, increasing the value of $\tilde U$ destroys localisation between left and right chains ($n_R\approx n_L\lesssim 0.5$).

\begin{figure}
\begin{center}
 \includegraphics[width=.48\textwidth]{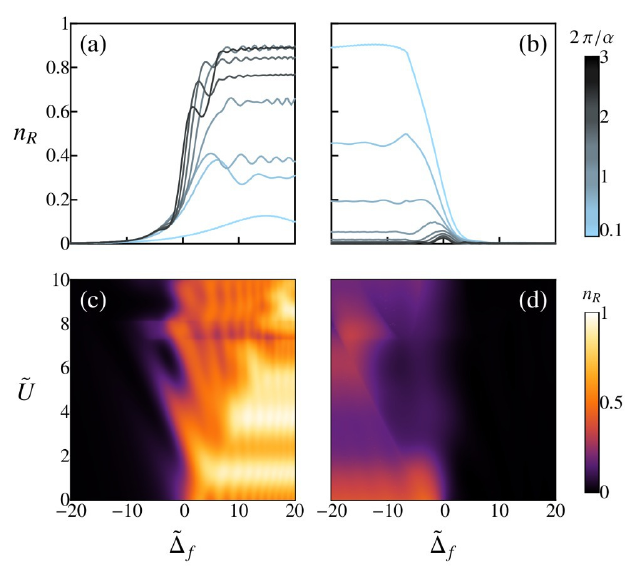}

\end{center}
\caption{(color on-line) Normalized transfer efficiency $n_R$ for ground-state (a) [inverse (b)] sweeps as a function of the final tilt $\tilde{\Delta}_f$, for $\tilde{U}=1.5$ and $2\pi/\alpha\in [0.1,3.0]$. Normalized transfer efficiency $n_R$ as a function of the final tilt $\tilde{\Delta}_f$, for $2\pi/\alpha=2.0$, [$2\pi/\alpha=1.0$] in the ground-state (c) [inverse (d)] case and $\tilde{U}\in [0,10]$. The number of particles  is $N=2$. Dimensionless parameters are: $n_R$, $\tilde{\Delta}_f$, $\tilde U$ and $\alpha$. }
\label{fig7}
\end{figure}

To compare the dynamical behavior between the ground-state and inverse sweeps we study, as in the experimental counterpart, the transferred population as a function of the final value of the tilt $\tilde \Delta_f$, for $N=2$, a fixed sweep rate $\alpha$ and fixed interaction $\tilde U$. As before, the initial state for the ground-state (inverse) sweep corresponds to the total particle population localized  in the bottom well with lower (higher) energy. However, it is important to point out that in this case the sweeps are carried out from $\tilde \Delta_i$ to $\tilde \Delta_f$, that is, the sweeps are not accomplished from $-|\tilde \Delta_i|$ to $|\tilde \Delta_i|$ as in section \ref{PD}. As initial condition for each sweep we consider $\tilde\Delta_i=20$ and $\tilde\Delta_i=-20$ for ground-state and inverse sweeps respectively.  Fig. \ref{fig7} corresponds to ground-state (a,c) [ inverse (b,d)] sweeps. In Fig. \ref{fig7} (a) [(b)] we plot $n_R$ for ground-state [inverse] sweeps as a function of $\tilde \Delta_f$ for $\tilde U=1.5$ and several values of $2\pi/\alpha$. In  Fig. \ref{fig7} (c) [(d)] we plot  $n_R$ for ground-state [inverse]  sweeps as a function of $\tilde \Delta_f$ for $2\pi/\alpha=2.0 [1.0]$  and $\tilde{U}\in[0,10]$.  As can be observed from these figures the ground state sweeps show that the normalized  population transferred increases from zero to a maximum $\max(n_R)<1$ for $2\pi/\alpha\lesssim1$ while the behavior for the inverse sweeps shows that the population transferred is partial, reaching a maximum value below one. The effect of increasing interaction for a fixed sweep rate, is to suppress the maximum population transfer. Comparison with the experimental results show qualitative agreement. See Fig. 3 of Ref.  \cite{IBloch}.

\section{Final remarks} 

We have studied the dynamical and stationary properties of an interacting Bose quantum fluid confined in two coupled one-dimensional chains. Such a system has been realized experimentally  \cite{IBloch} and represents a generalization of the single-particle Landau-Zener dynamics. The competition terms that determine the evolution in time from a given initial state depend on the inter particle interaction strength $U$, the intra-chain coupling energy $J^{\parallel}$ ($z$ direction), the inter-chain coupling energy $J^{\perp}$ ($x$ direction) and the time-dependent parameter characterizing the tilt between the two wells of each site in the chain, that is, the detunning parameter $\Delta$. 

To investigate the dynamics of the generalized LZ realization, we work in the decoupling approximation and the one-level band scheme, and propose a model lattice site Hamiltonian that represents two coupled infinite chains (\ref{Hmodel}). Our full many-body model is written in terms of the order parameters  $\psi_\nu$, $\nu \in \{L,R\}$ that account for the superfluid component of the particle population. We exhaustively explore the space of parameters $\{\tilde{U},\tilde{\Delta}\}$ to determine by variational means the order parameters $\psi_\nu$, $\nu \in \{L,R\}$ for the ground state energy  $\tilde E_0(\tilde U,\tilde \Delta,\psi_L, \psi_R)$. Then, we diagonalized the Hamiltonian to obtain the energy spectrum.  Concerning the stationary state properties, since the number of particles per lattice sites in the actual experiments is not too large, we were able to perform accurate numerical simulations of the system to fully characterize the entire energy spectrum. Such calculations lead to predict a richer spectrum with respect to that performed for a single dimer, that is, when the movement along $z$ axis is neglected \cite{IBloch}. We found that as $\tilde U$ grows in magnitude, a more complex structure of the energy spectrum with respect to the double-well system is developed in correlation with the suppression of the superfluid component in the system, thus promoting self-trapping behavior. In addition our model allow us to recover the single-particle LZ result for a single dimer.

Regarding the dynamical evolution of the LZ many body generalization, we concentrate in studying ground-state and inverse sweeps in order to compare the predictions of our model with the experimental realization. The ground-state and inverse sweeps dynamics account for a sweep from a state where all the particles initiate in the bottom well with lower energy (left well) and the counterpart where the filled sites are the ones with higher energy, respectively. In a phase diagram in terms of the interaction $\tilde U$ and the sweep rate $\alpha$ we condense the normalized final transfer of atoms. A totally different scenario with respect to the original single-particle Landau-Zener scheme was found for ground-state sweeps. That is, the transfer of particles does not vanish as the sweep rate tends to zero, on the contrary, when the sweep rate $\alpha$ takes smaller values, the total population initially placed on the left side is transferred almost completely to the right chain. A different situation occurs in the study of the transfer efficiency for inverse sweeps, this reveals consistent results with the single-particle Landau-Zener phenomena predictions. As the system evolves in time depending on the interaction, the system can transit different phases (superfluid or insulating) multiple times, see Fig. \ref{fig2}. We identified a a critical value of the interaction $\tilde U_c$, for which $\tilde{U} \gtrsim\tilde{U}_c$, the system is in a crossover from insulating states ($n_R\neq n_L$) with $\psi_{R,L}=0$ to a superfluid states ($n_R\approx n_L\lesssim 0.5$) with  $\psi_{R,L}\neq0$ . It is important to emphasize that all of our results are in qualitative agreement with the experimental realization. In addition, our analysis provides insight to the region where the interaction is non-perturvative.

{This work was partially supported by grants PAPIT  IN108812-2 and IN116110 DGAPA (UNAM). Computational facilities by DGSCA-UNAM are also acknowledged. We would like to thank V. Gal for useful discussions regarding the implementation of computation algorithms.}

\appendix*
\section{Single particle in a tilted double well}

 \begin{figure}
\begin{center}
 \includegraphics[width=.48\textwidth]{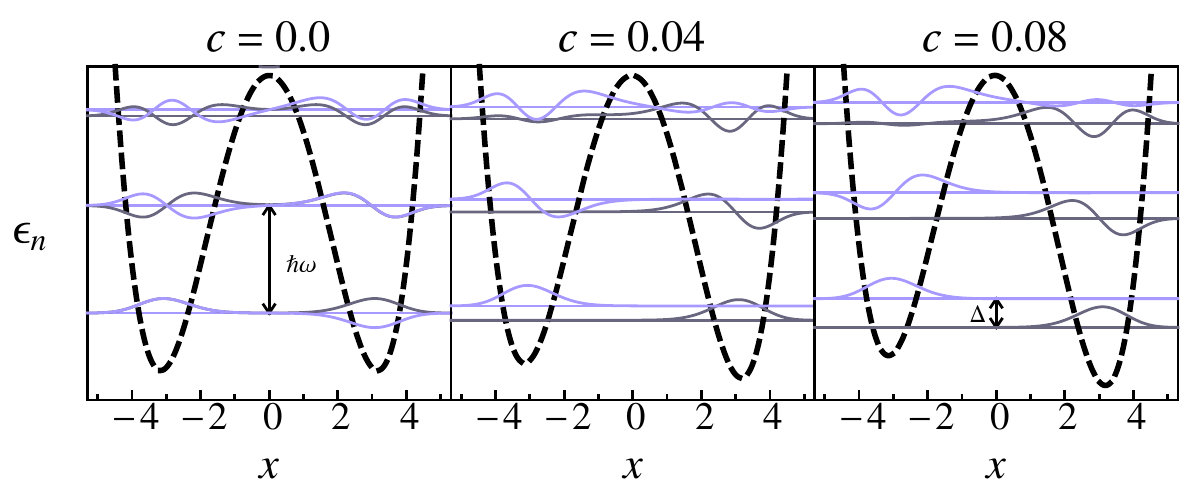}
\end{center}
\caption{(color on-line)
Energy spectrum of a single-particle in a double well potential.  The asymmetry is determined by the value of the parameter $c$ in (\ref{twell}). In these plots $c=0, 0.04$ and 0.08.}
\label{Af1}
\end{figure}

To establish appropriate bounds for the values of the parameters that determine the dynamics of the Bose gas in the coupled one-dimensional chains, that is, their practical use in relation to the proposed model Hamiltonian (\ref{Hmodel}), we analyze here the eigensystem of a single particle confined in an asymmetric double well. Although a previous study has provided quantitative criteria regarding the  energy level scheme where a many-particle system can be properly described for those potentials \cite{Dounas-Frazer}, the intrinsic nature of the Landau-Zener phenomena, as well as, the proposed model require to justify the plausibility of the chosen parameters. 

Let us consider the potential
\begin{equation}
V(x)=c x-x^2+d x^4,
\label{twell}
\end{equation}
where $d=0.05$ and the parameter $c$ incorporates the asymmetry. In Fig. \ref{Af1} we show the energy spectrum, as well as, the single-particle wave functions corresponding to different values of $c$. From this figure we observe several features. As expected, the energy levels are arranged in bands of two levels. The energy difference between bands becomes similar to the energy difference intra-bands as the asymmetry in the potential grows. That is, as the asymmetry in the double-well potential, represented by the parameter $\Delta=\alpha t$ changes with time $t$, the energy spacing between bands starts to match the energy difference among levels in the bands. This fact imposes a restriction on the possible values of the $\Delta$ parameter. In principle, one could think that the suitable range for values of $\Delta$ is severely limited. However, we recall that in our calculations all the parameters are referred to the tunnelling coupling parameter $J=J^{\parallel}=J^{\perp}$, which is related to the energy difference of the  two first energy levels at $t=0$ ($c=0$) as $J=(\epsilon_1-\epsilon_0)/2$  \cite{Paredes}, being $\epsilon_0$ and $\epsilon_1$ the ground state and the first excited state energies of the single particle. Therefore, by restricting ourselves to those values of $\tilde \Delta = \Delta/J \ << \hbar \omega/J$ and $\tilde U =U/J << \hbar \omega/J$, being $\hbar \omega\approx(\epsilon_3+\epsilon_2-\epsilon_1-\epsilon_0)/2$ the energy difference among bands, we can validate the one-band scheme to determine the dynamics of the proposed model Hamiltonian.  For the chosen double-well potential (\ref{twell}) we found $\hbar\omega/J\approx1.2\times10^5$
 and consequently $\Delta_{max}/J\approx1.2\times10^5$. Thus, for values of $\tilde \Delta < \Delta_{    max}/J$ the choice of parameters in the simulations and our model are well justified.


\begin{thebibliography}{99}
\bibitem{dt} The approximation of the time evolution is accurate to $O({\delta t}^2)$.
\bibitem{numerics2} The interval of study in $2\pi/\alpha$ and $\tilde U$ is $10/2^{9}$, that is, we explore a grid of size $2^9 \times 2^9$.
\end{thebibliography}
\end{document}